\documentclass[aps,prc,preprint,amsmath,amssymb,showpacs,preprintnumbers,superscriptaddress]{revtex4-1}
\usepackage{CJK}
\usepackage{graphicx}
\usepackage{dcolumn}
\usepackage{bm}

\usepackage{hyperref}

\begin{document}

\title{Magnetic rotations in $^{198}$Pb and $^{199}$Pb within covariant density functional theory with pairing correlations}
\author{Y. K. Wang}
\affiliation{State Key Laboratory of Nuclear Physics and Technology, School of Physics, Peking University, Beijing 100871, China}

\date{\today}
\begin{abstract}
  The observed shears bands in $^{198}$Pb and $^{199}$Pb are investigated within the framework of tilted axis cranking covariant density functional theory, and the separable pairing force is adopted to consider the pairing correlations.
  The energy spectra, total angular momenta, and transition probabilities agree well with the experimental data.
  The bandhead energy differences are reproduced satisfactorily without any \textit{ad hoc} renormalizations when the pairing correlations are taken into account.
  The angular momentum vectors are discussed in detail to analyze the pairing effects on magnetic rotations.
\end{abstract}
\maketitle

\date{today}

\section{Introduction}
The most common rotational bands in atomic nuclei are built on states with a substantial quadrupole deformation and characterized by strong quadrupole $(E2)$ transitions.
Such bands are usually well interpreted as the coherent collective rotation of many nucleons around an axis perpendicular to the symmetry axis of the deformed density distribution.
The study of these kinds of bands has been going on for several decades, and continues to generate much excitement due to the discoveries such as backbending~\cite{Johnson1971PhysicsLettersB605-608}, alignment phenomena~\cite{Stephens1972NuclearPhysicsA257-284,Banerjee1973NuclearPhysicsA366-382}, superdeformed rotational bands~\cite{Twin1986Phys.Rev.Lett.811--814}, etc.

In the early 1990s, however, the rotational-like sequences in weakly deformed neutron deficient Pb isotopes were observed~\cite{Clark1992PhysicsLettersB247-251}.
This new type of rotational bands has strong $M1$ and very weak $E2$ transitions.
Moreover, in this case, nuclei can rotate around an axis tilted with respect to the principal axes of the density distribution.
In order to distinguish this kind of rotation from the case in well-deformed nuclei (call electric rotation), the name ``magnetic rotation" was introduced in Ref.~\cite{Frauendorf1994}.
To data, more than 195 magnetic rotational bands have been observed in $A\sim 60, A\sim80, A\sim110$, $A\sim 140$, and $A\sim190$ mass regions~\cite{Meng2013Front.Phys.55}.

Theoretically, magnetic rotation was first explained by tilted axis cranking (TAC) model~\cite{Frauendorf1993Nucl.Phys.A259}.
In this model, it is relatively easy to construct classical vector diagrams showing the angular momentum composition, which is of great help in representing the structure of rotational bands.
The quality of the TAC approximation was examined in comparison with the quantum particle rotor model (PRM)~\cite{Frauendorf1996Z.Phys.A263}.
Based on the TAC model, many applications have been carried out in the framework of schematic Hamiltonians, such as the pairing plus quadrupole model~\cite{Frauendorf2001Rev.Mod.Phys.463}.

During the past decades, density functional theories~(DFTs)~have received wide attention owing to their successes in describing many nuclear phenomena in stable as well as in exotic nuclei~\cite{Ring1996Prog.Part.Nucl.Phys.193,Bender2003Rev.Mod.Phys.75,Meng2006Phys.Rev.C037303,Meng2006Prog.Part.Nucl.Phys470,Liang2015Phys.Rep.570,Meng2015J.Phys.G42}.
Considering the impressive achievements of DFTs, the extended version of TAC approaches, namely, the TAC approaches based on DFTs have been developed~\cite{Madokoro2000Phys.Rev.C061301,Peng2008Phys.Rev.C24313,Zhao2011Phys.Lett.B181,Meng2013Front.Phys.55,Meng2016Phys.Scr53008}.
These self-consistent methods are based on more realistic two-body interactions and, thus, can be used to investigate the nuclear rotational excitations on a more fundamental level with all important effects included, such as core polarization and nuclear currents~\cite{Olbratowski2004Phys.Rev.Lett.52501,Zhao2011Phys.Rev.Lett.122501}.
In particular, the tilted axis cranking covariant density functional theory (TAC-CDFT) provides a self-consistent description of currents and time-odd fields, and the included nuclear magnetism~\cite{Koepf1989Nucl.Phys.A61} plays an important role in the description of nuclear rotations~\cite{Konig1993Phys.Rev.Lett.3079,Afanasjev2000Phys.Rev.C31302,Afanasjev2010Phys.Rev.C34329,Liu2012Sci.ChinaPhys.Mech.Astron.2420}.
So far, the two-dimensional TAC-CDFT has been successfully used to describe the magnetic rotational bands~\cite{Zhao2011Phys.Lett.B181, Yu2012Phys.Rev.C024318, Peng2015Phys.Rev.C044329}, antimagnetic rotational bands~\cite{Zhao2011Phys.Rev.Lett.122501, Zhao2012Phys.Rev.C054310, Peng2015Phys.Rev.C044329}, linear $\alpha$ cluster bands~\cite{Zhao2015Phys.Rev.Lett022501}, etc. High predictive power has been demonstrated as well~\cite{Meng2013Front.Phys.55,Meng2016Phys.Scr53008}.

In previous investigations on nuclear rotational excitations within the framework of TAC-CDFT, pairing correlations are neglected.
The pairing effects on tilted axis rotational bands are not well understood until TAC-CDFT with a monopole pairing force is developed in Ref.~\cite{Zhao2015Phys.Rev.C034319}.
It is found that the description of energy spectra, especially the bandhead energy differences, can be significantly improved by the inclusion of pairing correlations.
Later on, TAC-CDFT with a separable pairing force is implemented to investigate the yrast sequences in $^{109}$Ag and good agreements with experimental data have been achieved~\cite{Wang2017Phys.Rev.C054324}.

In present work, the TAC-CDFT with a separable pairing force is used to study the classical magnetic rotational bands observed in $^{198}$Pb and $^{199}$Pb, and
the leftover problems in Ref.~\cite{Yu2012Phys.Rev.C024318}, namely, the renormalization of bandhead energies and the too low rotational frequencies for band crossings will be solved by considering the pairing correlations.

The paper is organized as follows: after sketching the formalism of the TAC-CDFT with the separable pairing force in Sec.~\ref{sec1}, the numerical details of the method are given in Sec.~\ref{sec2}.
In Sec.~\ref{sec3}, the calculated results with the corresponding data are compared and the pairing effects on the magnetic rotational bands in $^{198}$Pb and $^{199}$Pb are discussed.
Finally, a summary is given in Sec.~\ref{sec4}.

\section{Theoretical framework}\label{sec1}
For the consideration of pairing correlations in the framework of TAC-CDFT, as in Refs.~\cite{Zhao2015Phys.Rev.C034319, Wang2017Phys.Rev.C054324}, one needs to solve the following RHB equation,
\begin{equation}\label{eq1}
  \left(
  \begin{array}{cc}
    h-\bm{\omega}\cdot\bm{J}&\Delta\\
    -\Delta^\ast&-h^\ast+\bm{\omega}\cdot\bm{J}^\ast
  \end{array}
  \right)
  \left(
  \begin{array}{c}
    U_k\\
    V_k
  \end{array}
  \right) = E_k\left(
  \begin{array}{c}
    U_k\\
    V_k
  \end{array}\right),
\end{equation}
where $h$ is the single-nucleon Dirac equation
\begin{equation}
  h_D = \bm{\alpha}\cdot(\bm{p}-\bm{V})+\beta(m+S)+V
\end{equation}
minus the chemical potential $\lambda$, and $\Delta$ is the pairing field.
Here, the mean fields $S$ and $V^\mu$ as well as the pairing field $\Delta$ in Eq.~\eqref{eq1} are treated in a unified and self-consistent way.
The scalar and vector fields $S(\bm{r})$ and $V^\mu(\bm{r})$ are determined by
\begin{equation}
  \begin{split}
    S(\bm{r}) &= \alpha_S\rho_S+\beta_S\rho_S^2+\gamma_S\rho_S^3+\delta_S\Delta\rho_S, \\
    V^\mu(\bm{r}) &= \alpha_Vj^\mu_V+\gamma_V(j^\mu_V)^3+\delta_V\Delta j_V^\mu+\tau_3\alpha_{TV}j^\mu_{TV}\\
    &+\tau_3\delta_{TV}\Delta j^\mu_{TV}+eA^\mu,
  \end{split}
\end{equation}
with the densities and currents
\begin{align}
	\rho_S &= \sum_{k>0}{V}_k^\dag\gamma^0 V_k, \\
	j^\mu_V &= \sum_{k>0}\bar{V}_k\gamma^\mu V_k, \\
	j^\mu_{TV} &= \sum_{k>0}\bar{V}_k\gamma^\mu\vec{\tau}V_k,
\end{align}
and the electromagnetic field $eA^\mu$ with $e$ the electric charge unit vanishing for neutrons.

The matrix element of the pairing field $\Delta$ is
\begin{equation}
  \Delta_{ab} = \frac{1}{2}\sum_{c,d}\langle ab|V^{pp}|cd\rangle_a\kappa_{cd},
\end{equation}
where $V^{pp}$ is the pairing force, and the pairing tensor $\kappa = V^\ast U^T$ is determined by the quasiparticle (qp) wavefunctions. In the present work, the separable pairing force is adopted, which reads in the coordinate space,
\begin{equation}
  V^{pp}(\bm{r}_1,\bm{r}_2,\bm{r}_1',\bm{r}_2') = G\delta(\bm{R}-\bm{R}')P(\bm{r})P(\bm{r}')\frac{1}{2}(1-P^\sigma).
\end{equation}
Here, $\bm{R} = \frac{1}{2}(\bm{r}_1+\bm{r}_2)$ and $\bm{r}=\bm{r}_1-\bm{r}_2$ denote the center of mass and the relative coordinates respectively, and $P(\bm{r})$ has a Gaussian expression
\begin{equation}
  P(\bm{r}) = \frac{1}{(4\pi a^2)^{3/2}}e^{-r^2/4a^2}.
\end{equation}
The projector $\frac{1}{2}(1-P^\sigma)$ allows only the states with the total spin $S=0$.
The two parameters $G$ and $a$ have been determined in Ref.~\cite{Tian2009Phys.Lett.B44} by fitting to the density dependence of  pairing gaps at the Fermi surface for nuclear matter obtained with the Gogny forces.

By solving the Eq.~\eqref{eq1} iteratively, one can get the expectation values of the angular momentum, total energies, quadrupole moments, transition probabilities, etc.
The detailed formulas can be seen in Ref.~\cite{Wang2017Phys.Rev.C054324}.

\section{Numerical details}\label{sec2}
In present work, the observed magnetic rotational bands 1 and 3 in $^{198}$Pb as well as bands 1 and 2 in $^{199}$Pb are investigated.
Here, the same short-hand notation for the configurations are used as in Ref.~\cite{Yu2012Phys.Rev.C024318}: A, B, C, and D denote $i_{13/2}$ neutron holes with positive parity and E denotes a neutron hole in $(fp)$ shell with negative parity.
These four bands have the same proton configuration, i.e., $\pi(s^{-2}_{1/2}h_{9/2}i_{13/2})$, and the proton configuration is abbreviated by its spin number 11.
Using these short-hand notations, the configurations before back-bending for bands 1 and 3 in $^{198}$Pb are referred as AE11 and AB11, and the configurations after back-bending, caused by the two more aligned $i_{13/2}$ neutron holes, are referred as ABCE11 and ABCD11, respectively.
As for the bands 1 and 2 in $^{199}$Pb, the configurations before back-bending are referred as A11 and ABE11, and the configurations after back-bending, also caused by two more aligned $i_{13/2}$ neutron holes, are denoted by ABC11 and ABCDE11, respectively.

The TAC-CDFT with pairing correlations is applied to all of these four magnetic rotational bands.
The well-known density functional PC-PK1~\cite{Zhao2010Phys.Rev.C054319} is adopted in particle-hole channel.
The separable pairing force with $G=-738$ MeV fm$^3$ and $a = 0.636$ fm~\cite{Tian2009Phys.Lett.B44} is used in particle-particle channel.
The RHB equation~\eqref{eq1}~are solved with three dimensional harmonic oscillator basis in Cartesian coordinates with 12 major shells, which is the same as the case in Ref.~\cite{Yu2012Phys.Rev.C024318}.
In the present calculations, the method proposed in Ref.~\cite{Zhao2015Phys.Rev.C034319} is followed to trace and block the right qp orbits, and to keep the multi-qp configurations unchanged while solving Eq.~\eqref{eq1} iteratively with different $\lambda$ and rotational frequency $\omega$ values.

\section{Results and discussion}\label{sec3}
\begin{figure}[!htbp]
  \centerline{
  \includegraphics[width=0.8\textwidth]{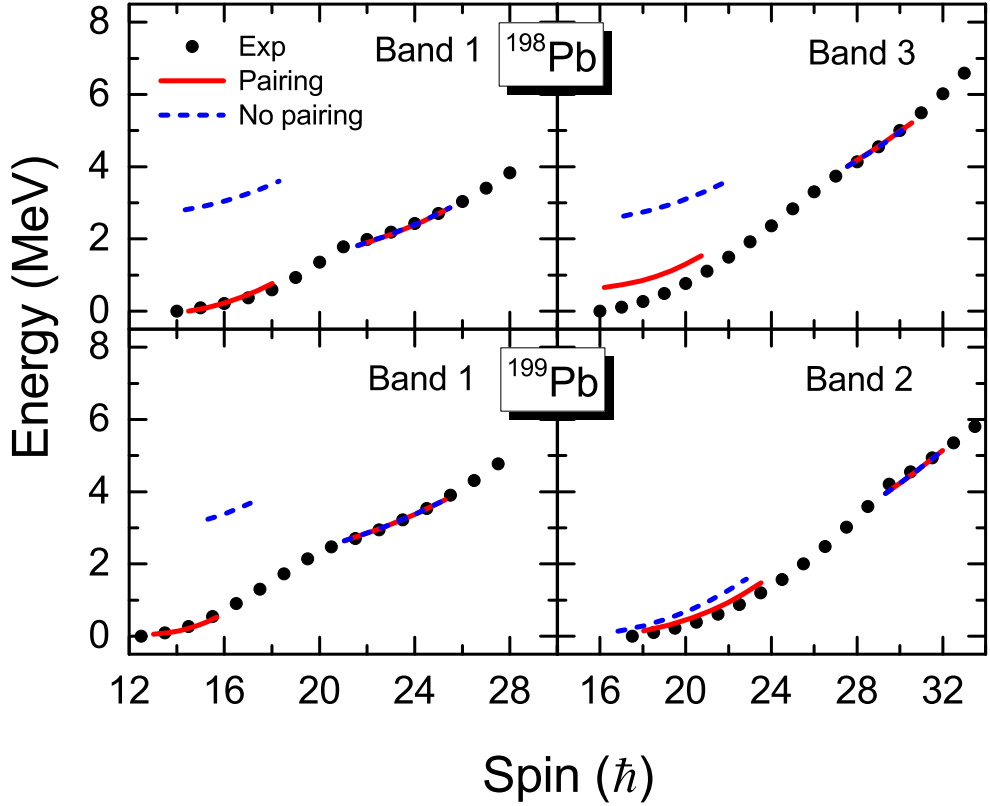}}
  \caption{(Color online)  Rotational excitation energies calculated by TAC-CDFT for the bands 1 and 3 in $^{198}$Pb (upper panels) and bands 1 and 2 in $^{199}$Pb (lower panels) as a function of total angular momentum, in comparison with the data~\cite{Goergen2001NuclearPhysicsA108-144,Lemaire-SemailB.1999Eur.Phys.J.AP257-267} (solid dots).
  The solid and dashed lines represent the calculated results with and without pairing correlations, respectively.}
  \label{fig1}
\end{figure}

The excitation energies for bands 1 and 3 in $^{198}$Pb as well as bands 1 and 2 in $^{199}$Pb calculated by TAC-CDFT with and without pairing correlations are shown in Fig.~\ref{fig1}, in comparison with the experimental data~\cite{Goergen2001NuclearPhysicsA108-144,Lemaire-SemailB.1999Eur.Phys.J.AP257-267}.
It is found that energy spectra for all of these four bands are well reproduced.
Especially, the energy differences before and after back-bending agree well with the experimental data without any artificial renormalizations of the bandhead after the inclusion of pairing correlations.
Meanwhile, one can determine the band crossing frequencies more accurately by considering the pairing correlations.
The pairing effects are significant for the bands based on configurations AE11 and AB11 in $^{198}$Pb, as well as A11 in $^{199}$Pb.
This leads to the remarkable reduction of excitation energies.
The effects of pairing correlations are relatively small for ABE11 in $^{199}$Pb due to the involved blocking neutron $i_{13/2}$ and $p_{1/2}$ orbits.
After the back-bending, caused by two more aligned $i_{13/2}$ neutron holes, the pairing effects vanish for all of the four bands and, thus, calculated results are same as the ones in Ref.~\cite{Yu2012Phys.Rev.C024318}.

\begin{figure}[!htbp]
  \centerline{
  \includegraphics[width=0.8\textwidth]{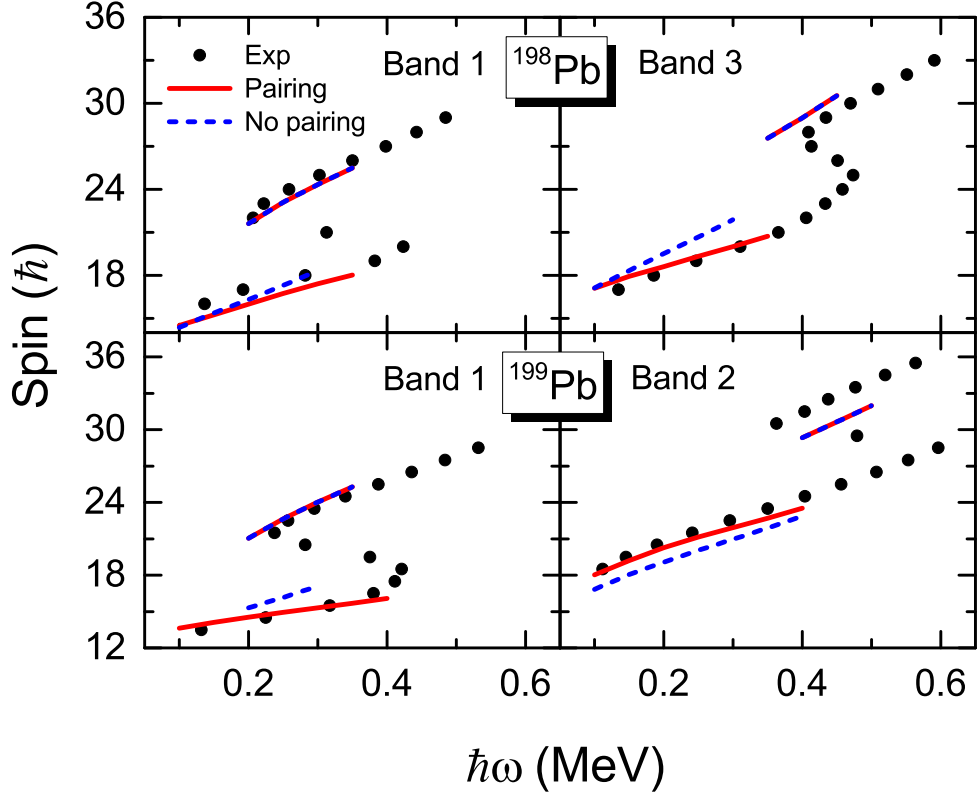}}
  \caption{(Color online) Total angular momenta calculated by TAC-CDFT for bands 1 and 3 in $^{198}$Pb (upper panels) and bands 1 and 2 in $^{199}$Pb (lower panels) as a function of rotational frequency, in comparison with the data~\cite{Goergen2001NuclearPhysicsA108-144,Lemaire-SemailB.1999Eur.Phys.J.AP257-267} (solid dots).
  The solid and dashed lines denote the calculated results with and without pairing correlations, respectively.}
  \label{fig2}
\end{figure}

The calculated total angular momenta as a function of rotational frequency are shown in Fig.~\ref{fig2}, in comparison with the experimental data~\cite{Goergen2001NuclearPhysicsA108-144,Lemaire-SemailB.1999Eur.Phys.J.AP257-267}.
The agreement with experiment is good with the inclusion of pairing correlations except for regions around the back-bendings, for example, in the spin region of $19-21\hbar$ for band 1 in $^{198}$Pb.
It is well known that the back-bending phenomenon is beyond the scope of a cranking calculation~\cite{Hamamoto1976Nucl.Phys.A15}.
In addition, from Fig.~\ref{fig2}, one can see that in the spin regions before back-bending of bands 1 and 3 in $^{198}$Pb as well as band 1 in $^{199}$Pb, the pairing effects on the total angular momenta are only visible in the high rotational frequency.
As for band 2 in $^{199}$Pb, the calculated total angular momenta with pairing correlations even larger than the cases without pairing at the same rotational frequency.
These phenomena are similar as demonstrated in Ref.~\cite{Zhao2015Phys.Rev.C034319}.

\begin{figure}[!htbp]
  \centerline{
  \includegraphics[width=0.8\textwidth]{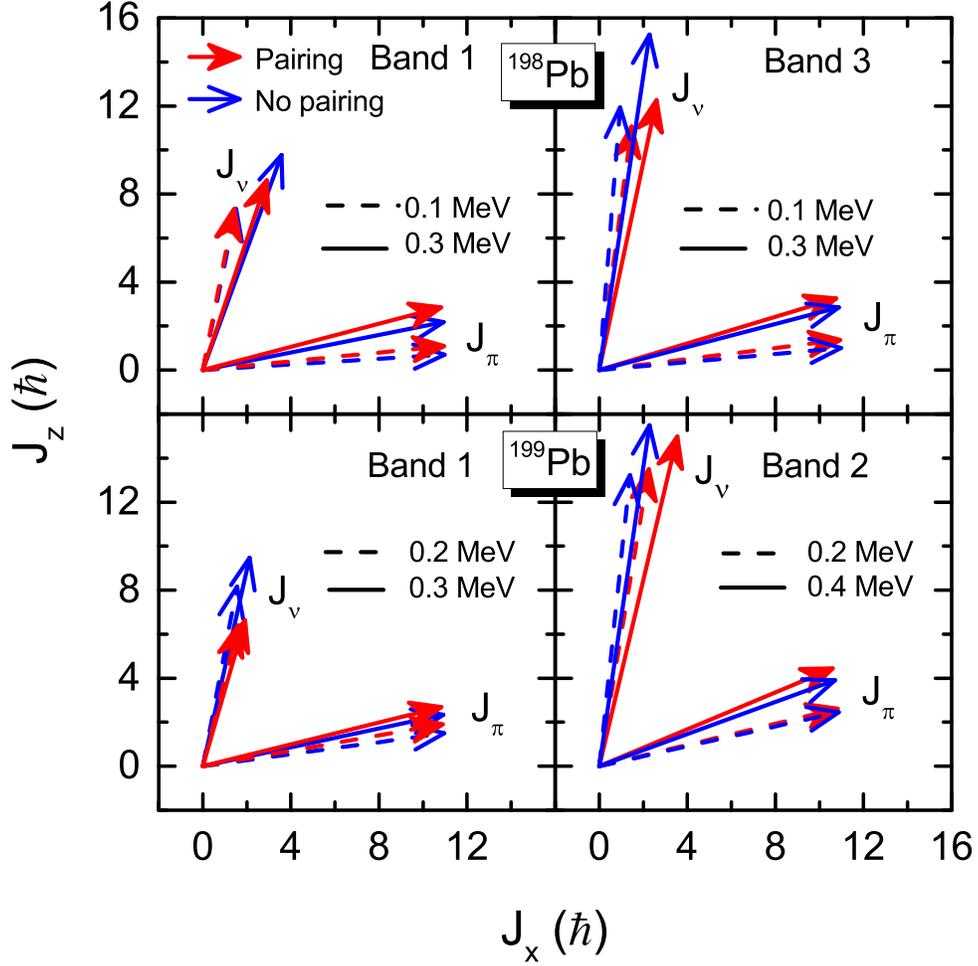}}
  \caption{(Color online) Neutron $\bm{J}_\nu$ and proton $\bm{J}_\pi$ angular momentum vectors for bands 1 and 3 in $^{198}$Pb (upper panels) and bands 1 and 2 in $^{199}$Pb (lower panels).
  Results with and without pairing correlations are distinguished by different types of arrows.}
  \label{fig3}
\end{figure}

In order to understand these distinctive features, the proton and neutron angular momentum vectors calculated by TAC-CDFT for the spin regions before back-bending are shown in Fig.~\ref{fig3}.
The results with and without pairing are distinguished by different types of arrows.
The solid and dashed lines are used to distinguish the results with different rotational frequencies.
As shown in Fig.~\ref{fig3}, for all the cases, the proton and neutron angular momentum vectors form the two blades of the shears.
As the increase of rotational frequency, these two blades move towards to each other and generate larger total angular momentum with the direction nearly unchanged.
In this way, the well-known ``shears mechanism" are clearly presented.

In present calculations, proton pairing energies for all of the bands vanish due to the large shell gaps in $^{198}$Pb and $^{199}$Pb with the proton number $Z = 82$.
Thus, the magnitudes of proton angular momentum vectors with and without pairing are almost equal to each other.
However, the magnitudes of neutron angular momentum vectors are reduced by pairing correlations.
It is noted that the angles between the two shears blades are also reduced by pairing correlations compared to the cases without pairing.
This leads to the increase of the total angular momentum.
Therefore, the pairing correlations have two competitive consequences, namely, reducing the amplitudes of proton and neutron angular momenta and closing the angles between them.

\begin{figure}[!htbp]
  \centerline{
  \includegraphics[width=0.8\textwidth]{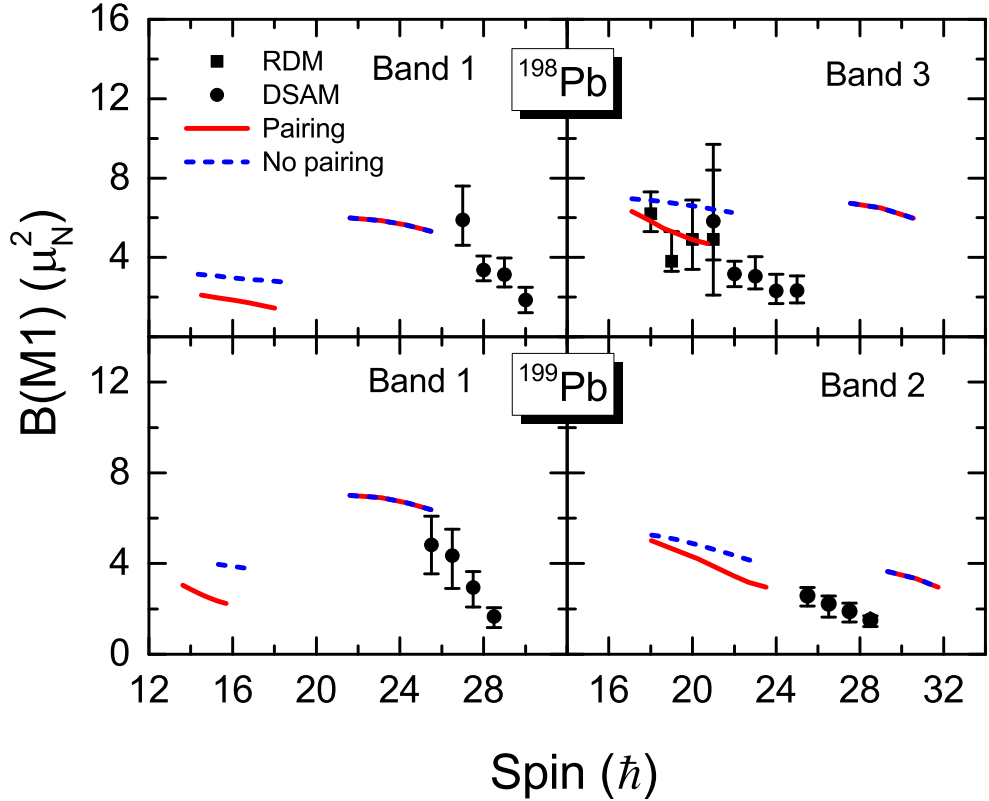}}
  \caption{(Color online) The calculated $B(M1)$ values for bands 1 and 3 in $^{198}$Pb (upper panels) and bands 1 and 2 in $^{199}$Pb (lower panels) as a function of the total angular momentum, in comparison with the data.
  Solid dots and squares denote experimental data from DSAM~\cite{Clark1997Phys.Rev.Lett.1868--1871} and FDM~\cite{Chmel2007Phys.Rev.C044309}, respectively.}
  \label{fig4}
\end{figure}

The transition probabilities $B(M1)$ associated with the four shears bands in $^{198}$Pb and $^{199}$Pb are also investigated, as shown in Fig.~\ref{fig4}.
The $B(M1)$ values are calculated with the semiclassical approximation from the magnetic moments~\cite{Frauendorf1997Nucl.Phys.A131}.
Here, the magnetic moments are derived from the relativistic electromagnetic current operator~\cite{Peng2008Phys.Rev.C24313}, and the Dirac effective mass scaled approximately to the nucleon mass by introducing a factor 0.58, similar to Refs.~\cite{Zhao2017PhysicsLettersB1-5, Wang2017Phys.Rev.C054324}.
A good agreement with the experiment is achieved by performing the TAC-CDFT calculations with pairing correlations.
For all of the magnetic rotational bands in $^{198}$Pb and $^{199}$Pb, the calculated $B(M1)$ values decrease with total angular momentum, which is a typical feature for the magnetic rotation.
For the spin regions before back-bending where pairing correlations play a role, the $B(M1)$ values are reduced compared to the results without pairing.
Meanwhile, the $B(M1)$ values decrease faster with total angular momentum which is associated with the fact that pairing can expedite the merging of the two shears blades as shown in Fig.~\ref{fig3}.

\begin{figure}[!htbp]
  \centerline{
  \includegraphics[width=0.8\textwidth]{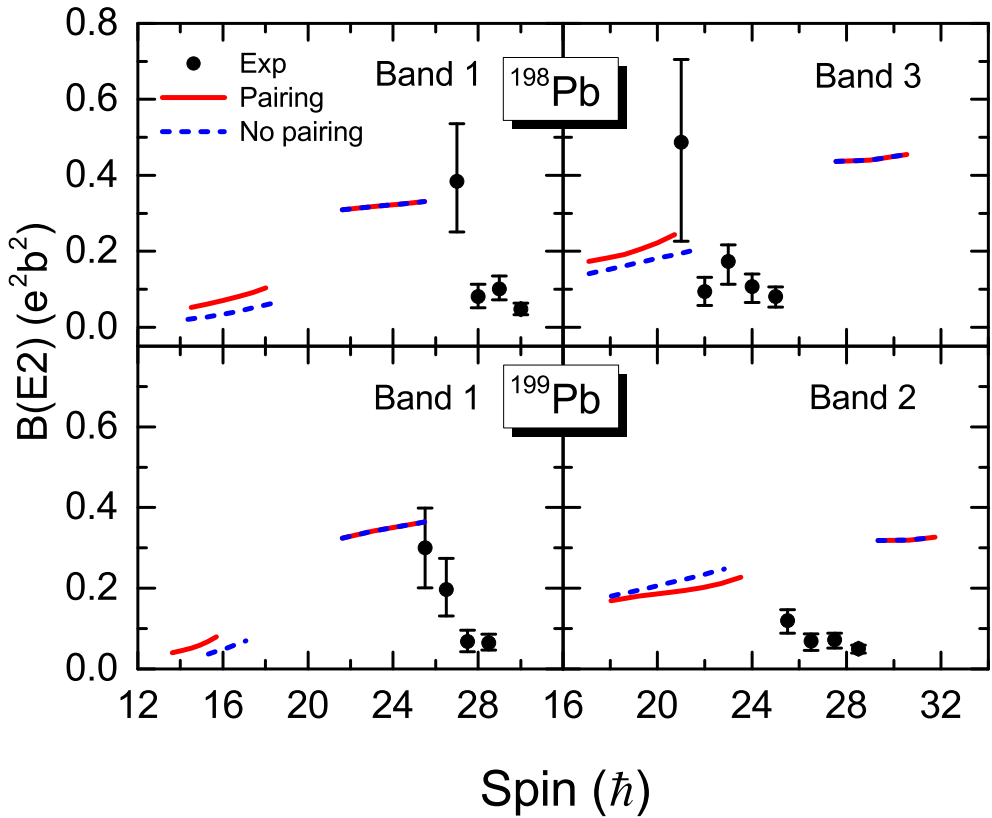}}
  \caption{(Color online)  The calculated $B(E2)$ values for bands 1 and 3 in $^{198}$Pb (upper panels) and bands 1 and 2 in $^{199}$Pb (lower panels) as a function of the total angular momentum, in comparison with the data~\cite{Clark1997Phys.Rev.Lett.1868--1871}.
  The solid and dashed lines denote the calculated results with and without pairing correlations, respectively.}
  \label{fig5}
\end{figure}

The $E2$ transition probabilities calculated by TAC-CDFT with and without pairing correlations are shown in Fig.~\ref{fig5}, in comparison with the experimental data.
The $B(E2)$ values in the TAC-CDFT calculations are in a reasonable agreement with experiment and keep roughly unchanged with total angular momentum.
This is consistent with the nearly constant quadrupole deformation in each configuration.
In contrast to the $M1$ transitions, pairing effects are marginal on $E2$ transitions.

\section{Summary}\label{sec4}
In summary, the TAC-CDFT with a separable pairing force is used to investigate the magnetic rotational bands observed in $^{198}$Pb and $^{199}$Pb.
The energy spectra, total angular momenta as well as transition probabilities are well reproduced after the pairing correlations are included.
The leftover problems, namely, the renormalization of bandheads and the too low rotational frequency for band crossings in Ref.~\cite{Yu2012Phys.Rev.C024318}, where the TAC-CDFT without pairing correlations are performed for the magnetic rotations in $^{198}$Pb and $^{199}$Pb, are solved by considering pairing correlations.
The conclusions about the impact of pairing on total angular momenta given in Ref.~\cite{Zhao2015Phys.Rev.C034319} hold true for present calculations in $^{198}$Pb and $^{199}$Pb.
In both calculations, pairing correlations can have two competitive consequences, namely, reducing the magnitude of proton and neutron angular momenta and expediting the merging of their directions.
These two counteracted consequences can, thus, influence the generation of total spin.

\begin{acknowledgments}
  The author thanks to J. Meng and P. W. Zhao for helpful discussions and careful readings of the manuscript.
  This work is supported in part by the Major State 973 Program of China (Grant No. 2013CB834400), the National Natural Science Foundation of China (Grants No. 11335002, No. 11375015, No. 11461141002, No. 11621131001).
\end{acknowledgments}

\newpage
%

\end{document}